\documentclass[prl,twocolumn,showpacs,amsmath,amssymb]{revtex4}
\usepackage{graphicx}
\usepackage{dcolumn}
\usepackage{bm}

\usepackage{color}

\usepackage{ulem}

\begin{document}

\title{Shape changes and large-amplitude collective dynamics in neutron-rich Cr isotopes
}

\author{Kenichi Yoshida and Nobuo Hinohara}
\affiliation{RIKEN Nishina Center for Accelerator-Based Science, Wako, Saitama 351-0198, Japan
}%

\date{\today}

\begin{abstract}
Shape-phase transition in neutron-rich Cr isotopes around $N=40$ is 
investigated by employing the collective Hamiltonian approach. 
The inertial functions for large-amplitude vibration and rotation
are evaluated 
by the local normal modes along the axial quadrupole collective coordinate  
using the Skyrme and pairing energy density functionals. 
The time-odd components of the mean fields are fully included in the derived masses.
The low-lying spectra obtained by requantizing the collective Hamiltonian 
show an excellent agreement with the recent experimental data. 
Gradual change from spherical to axially deformed shapes in between $N=34$ and 38 
is well described. 
\end{abstract}

\maketitle
How can we define the shape of quantum many-body systems? 
Atomic nuclei have a variety of equilibrium shapes, 
and show shape changes along the isotopic or isotonic chain. 
The mean-field approximation gives us an intuitive picture 
of the nuclear shape. 
However, we have to go beyond it to describe the shape-phase transition; 
the dynamical change of the mean-field potential associated with 
the large-amplitude collective motion. 

The self-consistent mean-field model employing the effective 
interaction or the nuclear energy-density-functional (EDF) method 
has successfully described the ground-state properties~\cite{ben03}. 
Recent advances in the computing capability
together with the highly-developed techniques in
the nuclear EDF method allow us to calculate
the ground-state properties of nuclei including deformation 
in the entire mass region of the nuclear chart~\cite{sto03}.

Magic number or shell closure is an essential concept in 
understanding the stability against the deformation. 
The subshell closure at 40 created by the gap between $1g_{9/2}$ and $2p_{1/2}, 1f_{5/2}$ 
orbitals has attracted much attention for several reasons~\cite{rei99}. 
The proton-rich $N=Z$ nucleus $^{80}$Zr lies in the center of the well-deformed $A\simeq 80$ 
region~\cite{naz85}. 
This is because a shell gap of 40 again appears in a deformed region, 
and the shell effect of protons and neutrons coherently stabilize the nucleus deformed. 
On the other hand, the existence of $N=40$ subshell closure is suggested 
experimentally for the neutron-rich nucleus $^{68}$Ni~\cite{bro95,sor02}.  
The strength of this subshell closure and its persistence for $Z<28$ 
determine the waiting point for the r-process nucleosynthesis at $^{64}$Cr, 
which is considered to be a progenitor of $^{64}$Ni~\cite{sor99}. 

The half-life measurement at CERN-ISOLDE have deduced that the neutron-rich 
$^{66}$Fe is deformed with a quadrupole deformation $\beta \sim 0.26$~\cite{han99}. 
Since the Cr isotopes lie at mid proton $1f_{7/2}$ shell, protons 
could additionally destabilize the nucleus in a spherical shape 
and favor deformation. 
Experimental evidences of the nuclear shape changes are related to
low-lying quadrupole collectivity, such as 
the small excitation energy of the $2^{+}_{1}$ state, 
the ratio of the excitation energy of 
the $2^{+}_{1}$ and $4^{+}_{1}$ states $R_{4/2}=E_{4+}/E_{2+}$, and 
the reduced transition probability $B(E2;2^{+}_{1}\rightarrow 0^{+}_{1})$, etc. 
The observed small excitation energy of the $2^{+}_{1}$ state 
in neutron-rich Cr isotopes 
indicates that the deformation develops toward $N=40$~\cite{sor03,bur05,zhu06,aoi09,gad10}.

Very recently, the large-scale shell-model calculation 
for the neutron-rich Cr isotopes has become available~\cite{len10}, 
where the Hamiltonian with dimension of $\sim 10^{10}$ is diagonalized. 
However, deducing the picture of deformation is difficult in the fully quantum-mechanical 
calculations. 

In this article, on the basis of the nuclear EDF method 
closely related to the mean-field approximation, 
we develop a new framework of the microscopic theory 
for the large-amplitude collective motion employing the quadrupole 
collective Hamiltonian approach for the axially-symmetric nuclei. 
The similar attempts of the EDF-based collective Hamiltonian starting from 
the Gogny interaction~\cite{lib99,gau09}, 
the relativistic Lagrangian~\cite{nik09,li10} 
and the Skyrme interaction~\cite{pro09} 
have been made for description of the large-amplitude collective motion. 
The cranking approximation, however, has been applied to calculate the inertial 
functions~\cite{lib99,gau09,nik09,li10,pro09}, 
and the time-odd components of the mean field remain largely unexplored 
except for a few attempts 
in the adiabatic time-dependent Hartree-Fock-Bogoliubov theories~\cite{dob81,gia80a,gia80b}. 
Our method includes the time-odd mean fields in the inertial functions. 
And this method is applied to the shape-phase transition in neutron-rich 
Cr isotopes around $N=40$.

We start from the $(1+2)$D
(vibration along the $\beta$ direction and rotations about the 
two axes which are perpendicular to the symmetry axis) 
quadrupole collective Hamiltonian;
\begin{equation}
\mathcal{H}_{\mathrm{coll}}=\dfrac{1}{2}\mathcal{M}_{\beta}(\beta)\dot{\beta}^{2}
+\dfrac{1}{2}\sum_{i=1}^{2}\mathcal{J}_{i}(\beta)\omega_{i}^{2}+V(\beta).
\label{cla_hami}
\end{equation}
With the Pauli's prescription, we requantize Eq.~$(\ref{cla_hami})$ 
and construct the collective Schr\"odinger equation;
\begin{equation}
\hat{H}_{\mathrm{coll}}\Psi_{\alpha IM}(\beta,\theta_{1},\theta_{2})=
E_{\alpha I}\Psi_{\alpha IM}(\beta,\theta_{1},\theta_{2}),
\label{hami}
\end{equation}
where 
$\Psi_{\alpha IM}(\beta,\theta_{1},\theta_{2})$ is the collective wave functions 
in the laboratory frame, $I$ is the angular momentum quantum number, 
$M$ is the $z-$component of $I$, and $E_{\alpha I}$ is the excitation energy. 
The collective wave functions are written 
in terms of wave functions in the 
body-fixed frame $\Phi_{\alpha IK}(\beta)$;
\begin{equation}
\Psi_{\alpha IM}(\beta,\Omega)=\sum_{K=0}^{I}\Phi_{\alpha IK}(\beta)
\langle \Omega|IMK\rangle=\Phi_{\alpha I}(\beta)\langle \Omega|IM0\rangle,
\end{equation}
with $\langle \Omega|IMK\rangle$ a superposition of the rotational wave functions. 

The collective Schr\"odinger equation in the intrinsic frame now reads
\begin{equation}
\left \{ 
\hat{T}_{\mathrm{vib}} + \dfrac{I(I+1)}{2\mathcal{J}(\beta)}+V(\beta)
\right \} 
\Phi_{\alpha I}(\beta)=E_{\alpha I}\Phi_{\alpha I}(\beta), \label{coll_Sch}
\end{equation}
where 
\begin{align}
\hat{T}_{\mathrm{vib}}=&
-\dfrac{1}{2M_{\beta}(\beta)}\dfrac{\partial^{2}}{\partial \beta^{2}} \notag \\
& +\dfrac{1}{2M_{\beta}(\beta)}
\left[
\dfrac{1}{2M_{\beta}(\beta)}\dfrac{\partial M_{\beta}(\beta)}{\partial \beta}
-\dfrac{1}{\mathcal{J}(\beta)}\dfrac{\partial \mathcal{J}(\beta)}{\partial \beta}
\right]
\dfrac{\partial}{\partial \beta},
\end{align}
and the vibrational wave function is normalized as
\begin{equation}
\int d\beta \Phi^{*}_{\alpha I}(\beta)\Phi_{\alpha^{\prime} I}(\beta)|G(\beta)|^{1/2}
=\delta_{\alpha \alpha^{\prime}}
\end{equation}
with the metric $|G(\beta)|=M_{\beta}(\beta)\mathcal{J}^{2}(\beta)$. 

The collective potential $V(\beta)$ is calculated by 
solving the constrained Hartree-Fock-Bogoliubov (CHFB) equation;
\begin{align}
&\delta \langle \phi(\beta)|\hat{H}_{\mathrm{CHFB}}|\phi(\beta)\rangle =0, \label{CHFB1}\\
&\hat{H}_{\mathrm{CHFB}}=\hat{H}-\sum_{\tau}\lambda^{\tau}\hat{N}^{\tau}
-\mu\hat{Q}_{20}. \label{CHFB2}
\end{align}

The microscopic Hamiltonian $\hat{H}$ is constructed from the
Skyrme and pairing EDFs. 
Following the discussion in Ref.~\cite{hin10}, 
the vibrational mass $M_{\beta}(\beta)$ is 
evaluated using the local normal mode as
\begin{align}
M_{\beta}(\beta) 
&= \dfrac{1}{\eta^{2}}\dfrac{\partial q_{\beta}}{\partial Q_{20}}\dfrac{\partial q_{\beta}}{\partial Q_{20}}, \\
\dfrac{\partial Q_{20}}{\partial q_{\beta}} 
&= \dfrac{\partial}{\partial q_{\beta}} \langle \phi(\beta)|\hat{Q}_{20}|\phi(\beta)\rangle
= \langle \phi(\beta)| [\hat{Q}_{20},\dfrac{\hat{P}^{\beta}}{i}]
|\phi(\beta)\rangle
\end{align}
by solving the local quasiparticle-random-phase approximation (LQRPA) 
equation on top of the CHFB state $|\phi(\beta) \rangle$ ;
\begin{align}
&\delta \langle \phi(\beta)|[\hat{H}_{\mathrm{CHFB}},\hat{Q}_{\nu}]-\dfrac{\hat{P}^{\nu}}{i} |\phi(\beta)\rangle=0, \label{QRPA1} \\
&\delta \langle \phi(\beta)|[\hat{H}_{\mathrm{CHFB}},\dfrac{\hat{P}^{\nu}}{i}]-\omega_{\nu}^{2}\hat{Q}_{\nu} |\phi(\beta)\rangle=0, \label{QRPA2}\\
&\langle \phi(\beta)|[\hat{Q}_{\mu},\dfrac{\hat{P}^{\nu}}{i}]|\phi(\beta)\rangle=\delta_{\mu}^{\nu}. \label{QRPA3}
\end{align}
Here, the isoscalar quadrupole moment operator $\hat{Q}_{20}$ and 
$\eta=\sqrt{\pi}/\sqrt{5}A\langle r^{2}\rangle$ are calculated at each state $|\phi(\beta)\rangle$. 
We write the collective coordinate as $q_{\beta}$. 
The rotational moment of inertia $\mathcal{J}(\beta)$ is evaluated by the 
LQRPA equation for collective rotation (extension of Thouless-Valatin equation to CHFB states~\cite{hin10}).

The reduced 
quadrupole matrix elements used for calculating 
the $E2$ transition probability and 
the spectroscopic quadrupole moment are evaluated as
\begin{align}
\langle \alpha I ||E2|| \alpha' I'\rangle =&
 \sqrt{(2I+1)(2I'+1)} (-)^I  \nonumber \\
& \begin{pmatrix}
I &2& I' \\ 0&0&0
\end{pmatrix}
\langle \alpha I |\hat{F}_{E2}| \alpha' I' \rangle, \\
\langle \alpha I|\hat{F}_{E2}|\alpha^{\prime}I^{\prime} \rangle
=& \int d\beta \Phi^{*}_{\alpha I}(\beta)F_{E2}(\beta)\Phi_{\alpha^{\prime} I^{\prime}}(\beta)
|G(\beta)|^{1/2},
\end{align}
where $F_{E2}(\beta)=\langle \phi(\beta)|\hat{F}_{E2}|\phi(\beta)\rangle$, 
and $\hat{F}_{E2}$ is the electric quadrupole moment operator.

We solve the CHFB plus LQRPA equations employing 
the extended procedures of Refs.~\cite{yos08,yos10}. 
To describe the nuclear deformation 
and the pairing correlations simultaneously with a good account of the continuum,
we solve the CHFB equations 
in the coordinate space using cylindrical coordinates
$\boldsymbol{r}=(\rho,z,\phi)$ with a mesh size of
$\Delta\rho=\Delta z=0.5$ fm and a box
boundary condition at $(\rho_{\mathrm{max}},z_{\mathrm{max}})=(12.25, 12.00)$ fm.
We assume axial and reflection symmetries for the CHFB states.
The quasiparticle (qp) states are truncated according to the qp
energy cutoff at $E_\alpha \leq 60$ MeV. 
We introduce the additional truncation for the LQRPA calculation,
in terms of the two-quasiparticle (2qp) energy as
$E_{\alpha}+E_{\beta} \leq 60$ MeV. 
In the present calculation, the LQRPA equations are solved 
on top of around 20 CHFB states. 
For each LQRPA calculation, we employ 256 CPUs and it takes 
about 185 CPU hours to calculate the inertial functions. 
Thus, for construction of the collective Hamiltonian, it takes 
about 3 700 CPU hours for each nucleus.

For the normal (particle-hole) part of the EDF,
we employ the SkM* functional~\cite{bar82}, 
and for the pairing energy, we adopt the volume-type pairing interaction as in Ref.~\cite{oba08}, 
where the deformation mechanism in neutron-rich Cr isotopes is investigated in detail.

\begin{figure}[t]
\begin{center}
\includegraphics[scale=0.68]{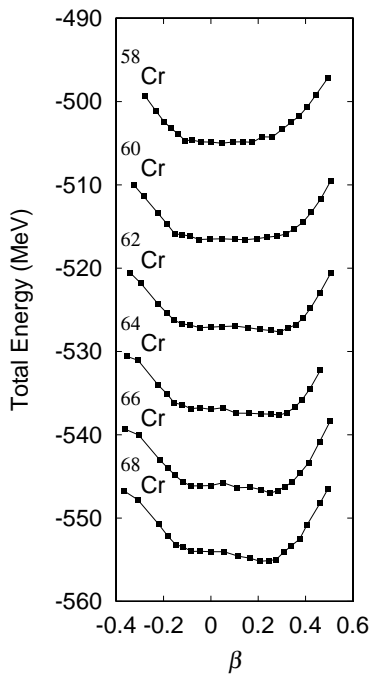}
\caption{Total energy curves of neutron-rich Cr isotopes calculated with SkM* and 
volume pairing interactions.
}
\label{deformation}
\end{center}
\end{figure}

Figure~\ref{deformation} shows the total energy curves
(collective potentials) of neutron-rich 
Cr isotopes under investigation. 
In $^{58}$Cr and $^{60}$Cr, the potentials are soft against the $\beta$ deformation 
even we get the HFB minima at $\beta=0$. 
Beyond $N=38$, the prolate minimum gradually develops up to $N=44$. 
We can see a shoulder or a shallow minimum at the oblately deformed region as well. 
It is noted that the systematic calculation predict 
$^{74}$Cr is spherical due to the magicity at $N=50$~\cite{sto03} 
using the same EDF as in the present calculation. 

\begin{figure}[t]
\begin{center}
\includegraphics[scale=0.47]{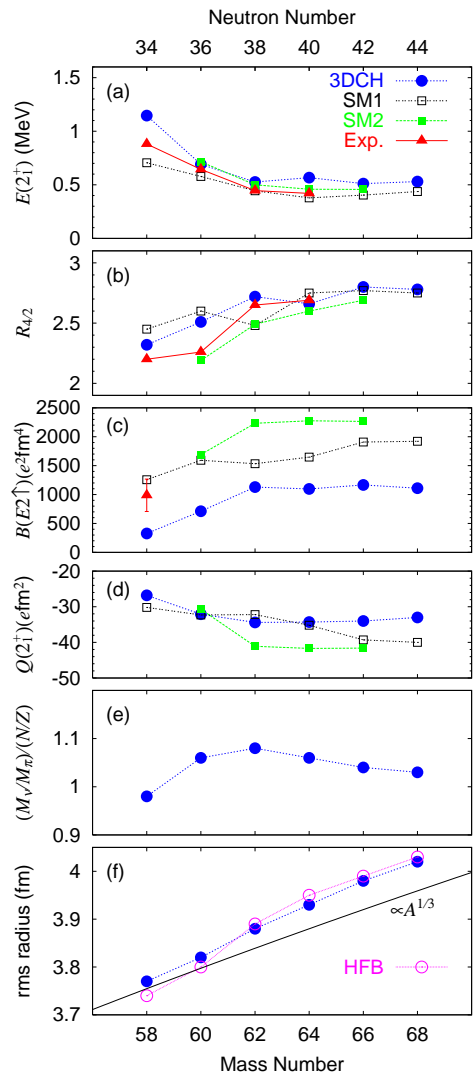}
\caption{(Color online) 
(a): Excitation energies of the $2^{+}_{1}$ state in Cr isotopes. 
(b): Ratios of the $E(4^{+}_{1})$ to $E(2^{+}_{1})$. 
(c): Reduced transition probabilities $B(E2;0^{+}_{1}\rightarrow 2^{+}_{1})$. 
(d): Spectroscopic quadrupole moment of the $2_{1}^{+}$ state. 
(e): Ratios of the transition matrix elements of neutrons to protons 
divided by the neutron to proton numbers. 
(f): Root-mean-square matter radii. 
Experimental data are taken from Refs.~\cite{bur05,zhu06,aoi09,gad10}. 
Results of the shell-model calculations (denoted as SM1~\cite{kan08} and SM2~\cite{len10}) 
are also included.
}
\label{yrast}
\end{center}
\end{figure}

Figure~\ref{yrast} shows the properties of low-lying states. 
When the neutron number increases from $^{58}$Cr, the excitation energy 
of the $2^{+}_{1}$ state drops toward $^{62}$Cr. 
The $R_{4/2}$ ratio and the transition 
probability $B(E2;0^{+}_{1}\rightarrow 2^{+}_{1})$ grow up at the same time. 
Although we overestimate $E(2^{+}_{1})$ at $N=34$, 
these features are consistent with the experimental results~\cite{zhu06,aoi09,gad10}, and 
it clearly shows the onset of deformation at $N \sim 38$.  
The experimental $B(E2)$ value is available only in $^{58}$Cr~\cite{bur05}. 
The present calculation underestimates the observation. 
Underestimation of the quadrupole collectivity of $^{58}$Cr 
probably attributes to the restriction of the collective coordinate 
to the $\beta$ and two rotational degree of freedom 
and/or to the EDF employed in the present calculation. 

Let us consider here the simplified case in solving Eq.~(\ref{coll_Sch}) with 
a harmonic oscillator potential and a constant vibrational mass;
\begin{equation}
V(\beta)=\dfrac{1}{2}C\beta^{2}, M_{\beta}(\beta)=B, 
\mathcal{J}(\beta)=3B\beta^{2}.
\end{equation}
We obtain $E(2^{+}_{1})=2.00$ MeV and $E(4^{+}_1)=4.26$ MeV, 
when we adopt $B=50$ MeV$^{-1}$ and $C=200$ MeV. 
The exact value for $E(2^{+}_1)$ is $\omega=\sqrt{C/B}=2$ MeV 
and $E(4^{+}_1)=2\omega=4$ MeV in the full five-dimensional collective Hamiltonian. 
This simple exercise implies 
that the present $(1+2)$D collective Hamiltonian 
can describe reasonably well the vibrational spectra of spherical nuclei. 
However, the $B(E2;2^{+}_1\to 0^{+}_1)$ value in the $(1+2)$D model 
is 3/5 times as large as an exact one.
Therefore, in (nearly) spherical nuclei, 
all the five quadrupole degrees of freedom should be treated on the same footing, 
and it remains as future work.

The results of the spherical shell-model calculations 
employing the pairing-plus-multipole forces with the monopole correction~\cite{kan08} 
and the LNPS interaction~\cite{len10} 
are also shown in Fig.~\ref{yrast}. 
Lowering of $E(2^{+}_{1})$ toward $N=38$ is consistent with the experiments 
and with the present calculation. 

We show in Fig.~\ref{yrast}(e) 
the ratios of the matrix elements of neutrons to protons $M_{\nu}/M_{\pi}$ 
divided by the neutron to proton numbers for 
the excitation to the $2^{+}_{1}$ state.
If the neutrons and protons coherently contribute to the excitation mode, 
$M_{\nu}/M_{\pi}$ may approach $N/Z$. 
We can see that the neutron excitation develops from $N=34$ to 38 
associated with the increase of the quadrupole collectivity.

Evolution of deformation is also seen in the matter radii as shown in Fig.~\ref{yrast}(f). 
For the reference, we show the $A^{1/3}$ dependence and the HFB results
by a solid line and open circles, respectively. 
In the deformed systems, the increase of radius is proportional to $\beta^{2}$. 
Thus, beyond $N=38$ the radius is constantly larger than the systematics of $A^{1/3}$. 
Furthermore, we can see that the calculated radius is larger than the HFB result 
in $^{58}$Cr and $^{60}$Cr associated with the large-amplitude fluctuation.

It is interesting to compare our results with those in Ref.~\cite{gau09}, 
where the collective Hamiltonian approach is also employed. 
In Ref.~\cite{gau09}, they employ the five-dimensional collective Hamiltonian,
in which both the $\beta$ and $\gamma$ degrees of freedom are included. 
And the D1S-Gogny EDF is used for the particle-hole and particle-particle parts of the energy density. 
The rotational mass is evaluated by the Inglis-Belyaev procedure, and 
the vibrational mass is calculated by use of the cranking approximation~\cite{lib99}.
In Ref.~\cite{gau09}, they obtained for $^{64}$Cr $E(2_{1}^{+})=0.82$ MeV, $R_{4/2}=2.2$, 
$Q(2_{1}^{+})=-34$ $e$fm$^{2}$, and $B(E2 \uparrow)=1515$ $e^{2}$fm$^{4}$.
The $B(E2)$ value and $Q(2^{+}_{1})$ are similar to our results, 
whereas $E(2^{+}_{1})$ and $E(4^{+}_{1})$ are rather larger. 
With the D1S-Gogny EDF, there is still a clear minimum at the spherical point, 
and a shallow local minimum around $\beta=0.3$~\cite{gau09}.
 
\begin{figure}[t]
\begin{center}
\includegraphics[scale=0.48]{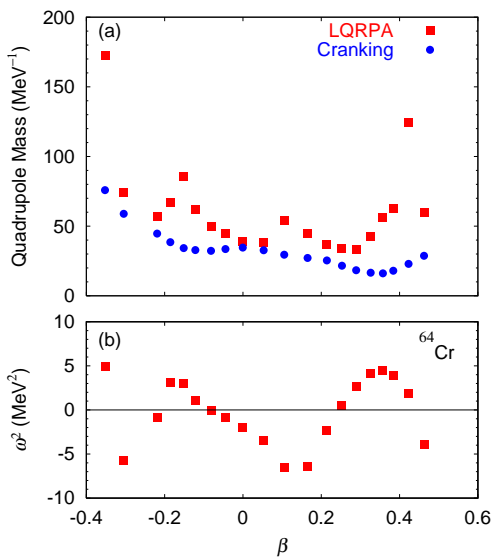}
\caption{(Color online) 
(a): Quadrupole masses calculated by the LQRPA and the cranking formula in $^{64}$Cr. 
(b): Excitation energies squared for the associated collective modes.
}
\label{64Cr_mass}
\end{center}
\end{figure}

In our approach, the time-odd components of the mean field are fully 
included for the calculation of the collective masses. 
Figure \ref{64Cr_mass}(a) shows the calculated vibrational masses 
along the collective coordinate. 
The collective masses calculated by use of the cranking approximation 
are also shown, where the time-odd components are neglected. 
The cranking masses show a smooth behavior as functions of the deformation $\beta$. 
This is because the cranking mass is evaluated by combination of the sum-rule values. 
In contrast to the cranking masses, the LQRPA masses strongly depend on 
the deformation, microscopic structure of the collective mode. 
And the LQRPA masses are larger than the cranking masses. 
In Fig.~\ref{64Cr_mass}(b), the excitation energies squared $\omega_{i}^{2}$ of the collective modes 
are shown. 
This quantity represents the curvature of 
the collective potential 
as seen in Eq.~(\ref{QRPA2}). 
The two-humped structure is associated with the existence of the two minima 
in the collective potential of $^{64}$Cr.

\begin{figure}[t]
\begin{center}
\includegraphics[scale=0.5]{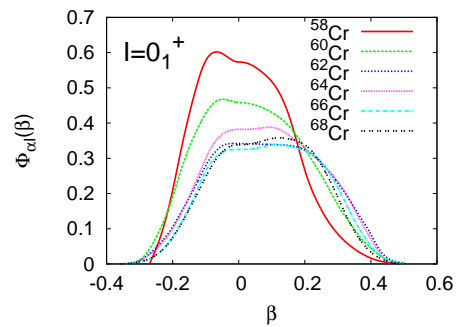}
\caption{(Color online)
Collective wave functions of the $I^{\pi}=0^{+}_{1}$ state in Cr isotopes.
}
\label{WF}
\end{center}
\end{figure}

Finally, we discuss the shape evolution of the ground state. 
Figure~\ref{WF} shows the vibrational wave functions of the $0^{+}_{1}$ state. 
In $^{58}$Cr, the wave function is distributed
around $\beta=0$. 
But the wave functions is not sharply localized 
at the spherical minimum of the potential, 
which reflects the softness of
the collective potential 
against the deformation. 
When two neutrons are added to $^{58}$Cr, broadening of the wave function can be seen. 
When two more neutrons are added, the wave function moves toward a prolately deformed 
minimum with a large spreading.
From this figure, we can say that $^{60}$Cr is located close to the critical 
point of the shape-phase transition in neuron-rich Cr isotopes and 
the large-amplitude dynamics plays a dominant role.

In summary, we have developed a new framework of the microscopic model 
for the large-amplitude collective motion based on the nuclear EDF method. 
The collective masses and the potential appearing in the quadrupole collective Hamiltonian 
are evaluated employing the constrained HFB and local QRPA approach, 
where the time-odd components of the mean field are fully taken into account. 
The microscopic Hamiltonian is constructed from 
the Skyrme and the pairing EDFs. 

We applied this new framework to the shape-phase transition in neutron-rich Cr isotopes 
around $N=40$. The present calculation gives consistent results for 
the low-lying excited states with the observations and the other theoretical approaches.  
Investigating the collective wave functions, we reach a conclusion that 
$^{60}$Cr is located close to the critical point of the shape phase transition, 
and the onset of deformation takes place at $N=38$. 
The large-amplitude dynamics plays a dominant role in the shape changes in 
neutron-rich Cr isotopes.

As a future work, it would be interesting to compare 
the present framework with the generator-coordinate method using the 
$\beta$-constrained HFB states. 
And the systematic investigation of the properties of low-lying collective states 
such as $E(2_{1}^{+})$, $R_{4/2}$ and $B(E2)$ in the entire region of nuclear chart 
is quite challenging both in nuclear structure physics and in computational science 
as a large-scale calculation employing the massively-parallel supercomputers.

The authors are supported by the Special Postdoctoral 
Researcher Program of RIKEN. 
The numerical calculations were performed 
on RIKEN Integrated Cluster of Clusters (RICC) and T2K-Tsukuba.

\end{document}